\documentclass[aps, prl, floatfix, reprint, superscriptaddress,
showpacs]{revtex4-1}
\usepackage{graphicx}
\draft 
\begin{document}
\title{Femtosecond intrapulse evolution of the magneto-optic Kerr effect 
\\in magnetoplasmonic crystals} 


\author{M. R. Shcherbakov}
\affiliation{Faculty of Physics, Lomonosov Moscow State University, Moscow 119991, Russia}
\affiliation{Samsung R\&D Institute Russia, Moscow, Russia}
\author{P. P. Vabishchevich}
\affiliation{Faculty of Physics, Lomonosov Moscow State University, Moscow 119991, Russia}
\author{A. Yu. Frolov}
\affiliation{Faculty of Physics, Lomonosov Moscow State University, Moscow 119991, Russia}
\author{T. V. Dolgova}
\affiliation{Faculty of Physics, Lomonosov Moscow State University, Moscow 119991, Russia}
\author{A. A. Fedyanin}
\email{fedyanin@nanolab.phys.msu.ru}
\affiliation{Faculty of Physics, Lomonosov Moscow State University, Moscow 119991, Russia}

\date{\today}
\begin{abstract} 
In magnetoplasmonics, it is possible to tailor the magneto-optical properties of nanostructures by exciting surface plasmon polaritons (SPPs). Thus far, magnetoplasmonic effects have been considered static. Here, we describe ultrafast manifestations of magnetoplasmonics by observing the non-trivial evolution of the transverse magneto-optic Kerr effect within 45-fs pulses reflected from an iron-based magnetoplasmonic crystal. The effect occurs for resonant SPP excitations, displays opposite time derivative signs for different slopes of the resonance, and is explained with the magnetization-dependent dispersion relation of SPPs.
\end{abstract}
\maketitle 

Since its establishment by Michael Faraday in 1845 \cite{Faraday}, magneto-optics has found numerous applications in science and technology, including methods like magnetic circular dichroism and magneto-optical microscopy, and devices such as magneto-optical isolators and memory. However, in the overwhelming majority of experiments, magneto-optic effects are measured with continuous-wave (CW) light sources.  It was only in 1996 that magneto-optics resorted to subpicosecond scale \cite{Beaurepaire}, when ultra-short laser pulses were demonstrated to affect the magnetic moments by either heating them or directing them with the pulse's magnetic field (see Ref.\cite{Kirilyuk} for a review). As femtosecond pulse sources were developed, immense possibilities in magnetic information recording and readout  were attained \cite{Stanciu}. However, a considerable downside of this approach lies in the requirement for intrinsic changes to be introduced into the medium by a high-power laser source.

To pursue shorter timescales for light-matter interactions, short-lived solid-state excitations, such as polaritons, can be utilized. Surface plasmon polaritons (SPPs), namely, electromagnetic waves that are bound to the free-electron plasma of a metal, are short-lived excitations that have durations of as much as several hundred femtoseconds \cite{Ropers, Vengurlekar, Vabishchevich}.  The interaction of a femtosecond laser pulse with plasmonic nanostructures has recently emerged as a topic for research \cite{Utikal, Stockman, Samson, Rokitski}, thus enabling, for example, laser pulse amplitude \cite{Brown} and polarization shaping \cite{Shcherbakov, Tok} with plasmonic media. 
Furthermore, it has been demonstrated that external quasistatic magnetic fields can be used to control the dispersion of SPPs in magnetic media \cite{Chiu}. However, despite the immense number of studies on magnetoplasmonics that have emerged during the last three decades \cite{Ferguson, Safarov, Hermann, Temnov, Grunin, GonzalezDiaz, Belotelov, FerreiroVila, Bonanni, Chetvertukhin}, magnetoplasmonic effects have thus far been considered static.

In this Letter, we experimentally demonstrate manifestations of a time-dependent transverse magneto-optic Kerr effect (TMOKE) within 45-fs laser pulses reflected from a one-dimensional iron-based magnetoplasmonic crystal. We show that exciting SPPs with magnetization-dependent dispersion enables control over the shape of the reflected pulse. The TMOKE evolution is demonstrated to have either a positive or negative time derivative, depending on the spectral position of the incident pulse's carrier wavelength, $\lambda_c$, with respect to the SPP resonance wavelength. Proper justification is given for this effect within the Lorentzian spectral line shape approach.

\begin{figure}[b!]
\includegraphics[width=0.95\columnwidth]{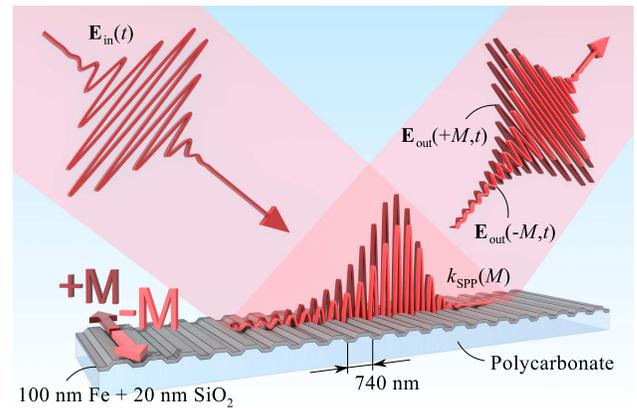}
\caption{Illustration of the ultrafast time-dependent TMOKE. The incident pulse is transformed into an SPP wave that has the dispersion law depending on the magnetization direction of the sample. The influence of SPPs is stronger at later time moments. Consequently, the reflected pulse profile depends on the sample magnetization, thereby yielding an intra-pulse time-dependent TMOKE.}\label{fig1}
\end{figure}

The schematic for creating appropriate conditions for observing an intra-pulse time-dependent TMOKE is illustrated in Fig.\,\ref{fig1}. The femtosecond pulse excites an SPP in a one-dimensional iron grating, the so-called magnetoplasmonic crystal. The SPP has a magnetization-dependent dispersion relation \cite{Grunin}:
\begin{equation}
k_{\mbox{\tiny SPP}}(M)=\frac{\omega}{c}\sqrt{\frac{\varepsilon}{\varepsilon+1}}\left[1+\alpha g(M)\right],\label{SPP}
\end{equation}
where $\varepsilon$ is the dielectric permittivity of iron, $\alpha=[\sqrt{-\varepsilon}(1-\varepsilon^2)]^{-1}$, and $g$ is the absolute value of the gyration vector of iron. The lifetime of the SPP is limited by radiative and dissipative losses to values of no greater than 1\,ps. Therefore, the resulting reflected pulse is perturbed with respect to the initial shape and contains information about the SPP near the end of the pulse, as depicted by the elongated tail of the reflected pulse in the schematic. The end of the pulse, therefore, is more sensitive to enhanced magnetoplasmonic effects; hence, one should expect an increased TMOKE near the end of the pulse.

\begin{figure}[t]
\includegraphics[width=\columnwidth]{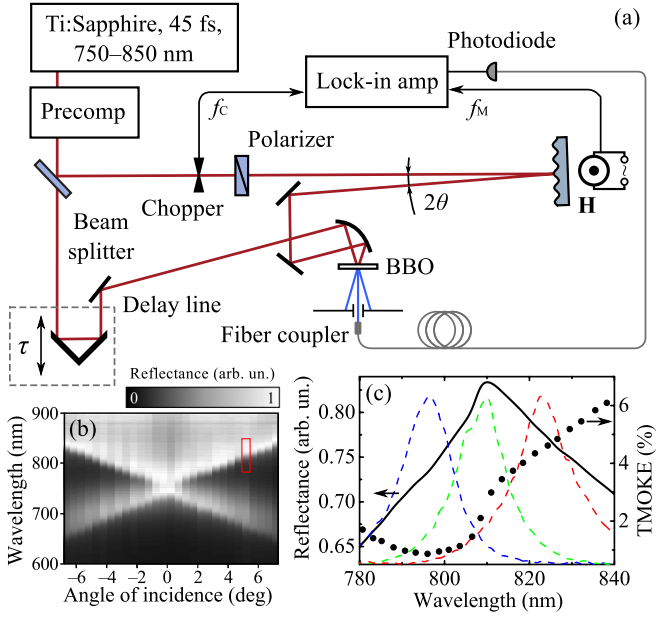}
\caption{(a) Experimental setup for observation of the femtosecond evolution of TMOKE. (b) Reflection spectrum of the $p$-polarized light from the sample as a function of the angle of incidence. The rectangle indicates the angle of incidence and the spectral range chosen for the experiment. (c) Reflectance and static TMOKE spectra as measured with a Ti:sapphire laser in the CW regime. The dashed curves denote the spectra of the femtosecond laser pulses with carrier wavelengths $\lambda_c=$ 795, 808, and 822\,nm in blue, green, and red, respectively.}\label{fig2}
\end{figure}

We briefly describe our experimental method to observe the non-trivial evolution of the TMOKE within femtosecond laser pulses. For measuring static TMOKEs, a quasi-monochromatic CW light source is routinely used. The changes in the sample's reflectance upon magnetization $M$ are monitored:
\begin{equation}
\delta(M)=\frac{R(M)-R(-M)}{R(0)},
\end{equation}
 where $R(M)$ is the reflectance of the sample as a function of the magnetization vector that lies in the plane of the sample perpendicular to the plane of incidence. If a pulsed source is used, however, it is possible to measure $\delta$ as a function of time within the pulse. The time-dependent TMOKE $\delta(t)$ is defined as
 \begin{equation}
 \delta(M,t)=\frac{I(M,t)-I(-M,t)}{I(0,t)},\label{delta}
 \end{equation}
 where $I(t)$ is the envelope function of the pulse. To obtain information about $\delta(t)$, the correlation function (CF) measurement setup depicted in Fig.\,\ref{fig2} was used. A train of 45-fs laser pulses from a Coherent Micra Ti:sapphire oscillator was pre-compressed at a chirped-mirror assembly to account for dispersive optics in the setup. The pulses were split into two beams, where one was the signal beam that contained the sample and the other was a gate beam with a 3-fs step delay line. Two electrically serial electromagnets placed around the sample were used to apply a quasistatic magnetic field in the TMOKE configuration, i.e., in the plane of the sample perpendicular to the plane of incidence. The magnets were driven by an AC source at a frequency of $f_M=117$\,Hz and provided approximately 30\,mT of magnetic flux density, which was sufficient to saturate the magnetization of the iron in the sample. Placing the non-magnetic sample holder and the magnets in an iron housing was found to prevent the other optomechanical components from mechanically oscillating at the frequency of the external magnetic field. The pulses in both beams were brought together by a parabolic mirror in a 100-$\mu$m-thick beta-barium borate (BBO) crystal that produced sum-frequency radiation aimed toward the core of a multi-mode optical fiber. The sum-frequency signal represents the second-order intensity correlation function of the intensity profiles of the two beams,
\begin{equation}
I_{\mbox{\tiny CF}}(M,\tau)=\int\limits_{-\infty}^{\infty}\left[I_{\mbox{\tiny sig}}(t)(1+\delta(M,t)/2)\right]I_{\mbox{\tiny gate}}(t-\tau)dt,
\end{equation}
and is therefore inherently dependent on the $\delta(M,t)$ function of the signal pulse. To avoid the magnetic field impact on the detector, a 2-m long fiber transferred the signal to a distant Si photodiode coupled to two lock-in amplifiers. Two signal values were measured: the first was the photocurrent amplitude $I_{\mbox{\tiny CF},f_{\mbox{\tiny C}}}$ at the frequency of the optical chopper, $f_C=420$\,Hz, which, in the absence of an external magnetic field, gives a value of
\begin{equation}
I_{\mbox{\tiny CF},f_{\mbox{\tiny C}}}(\tau)\propto\int\limits_{-\infty}^{\infty}I_{\mbox{\tiny sig}}(t)I_{\mbox{\tiny gate}}(t-\tau)dt.
\end{equation}
The second signal was the photocurrent amplitude $I_{\mbox{\tiny CF},f_{\mbox{\tiny M}}}$ at the frequency of the external magnetic field, which gives a value of
\begin{equation}
I_{\mbox{\tiny CF},f_{\mbox{\tiny M}}}(M,\tau)\propto\int\limits_{-\infty}^{\infty}\delta(M,t)I_{\mbox{\tiny sig}}(t)I_{\mbox{\tiny gate}}(t-\tau)dt.
\end{equation}
Finally, the ratio of these signals gives the value
\begin{equation}
\Delta(M,\tau)=
\frac{I_{\mbox{\tiny CF},f_{\mbox{\tiny M}}}}
{I_{\mbox{\tiny CF},f_{\mbox{\tiny C}}}}
 = 
 \frac{I_{\mbox{\tiny CF}}(M,\tau)-I_{\mbox{\tiny CF}}(-M,\tau)}{I_{\mbox{\tiny CF}}(\tau,0)},\label{Delta}
\end{equation}
which is a characteristic value of the time-dependent TMOKE signal. Although $\Delta(M,\tau)$ is a value that refers to $\delta(M,t)$ in an indirect fashion, its non-trivial dependence on $\tau$ indicates the non-trivial dynamics of the TMOKE. Otherwise, if $\delta(M,t)$ is time-independent,  $\Delta$ exactly equals the static TMOKE value.

Measurements of $\Delta(M,\tau)$ were performed for a one-dimensional magnetoplasmonic crystal based on a commercially available digital versatile disk polycarbonate template that had periodic corrugations with a depth of approximately 50\,nm and a period of 750$\pm$10\,nm. The dielectric template was covered by a 100-nm layer of polycrystalline iron deposited by magnetron sputtering and protected by a 20-nm-thick silica layer from the top. Angular-dependent reflectance spectroscopy indicated two branches of SPP modes as shown in Fig.\ref{fig2}(b). Based on the spectroscopy results, the angle of incidence of the laser radiation on the sample was chosen to be $\theta=5^\circ$. The particular angle of incidence was chosen intentionally (a) such that the reflectance maximum within one of the SPP branches was situated in the spectral vicinity of the laser radiation spectrum (see Fig.\ref{fig2}(c)) and (b) to eliminate the TMOKE caused by the iron surface itself, which scales as $\sin^22\theta$ \cite{Buschow}. Prior to the time-dependent measurements, static TMOKE spectroscopy was performed. First, significant enhancement of the TMOKE is shown in Fig.\,\ref{fig2}(c) compared with the value measured for a plain iron film, $\delta_{\mbox{\tiny Fe}}\approx0.1\%$, at the same angle of incidence. Strong wavelength dependence was observed in the TMOKE spectrum in the vicinity of the SPP resonance. This feature is connected to the magnetization-dependent dispersion relation of the SPP given in Eq.(\ref{SPP}). We will see below that Eq.(\ref{SPP}) also explains the time evolution of the TMOKE within the pulse.

\begin{figure}[t!]
\includegraphics[width=\columnwidth]{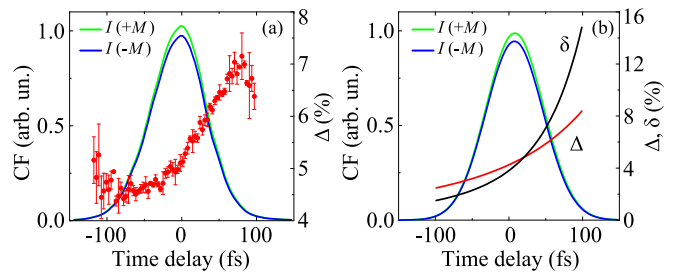}
\caption{(a) CFs measured for the $p$-polarized pulses at $\lambda_c=816$\,nm for an in-plane external magnetic field in the forward (green curve) and backward (blue curve) directions. The time dependence of the TMOKE is represented by the measured $\Delta(\tau)=[I(+M)-I(-M)]/I(0)$. (b) Modeling results for the corresponding experimental data using Eqs.(\ref{intensity},\ref{Eout}) (see the text for the parameter values). The time-dependent TMOKE, $\delta(t)$, is defined by Eq.(\ref{delta}); its CF counterpart, $\Delta(\tau)$, is defined by Eq.(\ref{Delta}).}\label{fig3}
\end{figure}

The effect of the magnetic field on the reflected pulse profile was also observed. Fig.\ref{fig3}(a) shows two CFs measured for the $p$-polarized pulses at $\lambda_c=816$\,nm with the magnetic field applied in one direction ($+M$, green curve) and also in the opposite direction ($-M$, blue curve). Manifestations of the TMOKE can be directly observed. Direct evidence of non-trivial TMOKE evolution is depicted in Fig.\ref{fig3}(a), with the measured $\Delta(\tau)$ function depicted with red dots and error bars. An overall increase in TMOKE is observed as the pulse evolves; this increase occurs because the end of the pulse contains more information about the SPP than the beginning of the pulse because the SPP has a finite decay time that is comparable to the pulse duration. The SPP causes TMOKE enhancement in the static case; therefore, the tail of the pulse provides a greater TMOKE than the beginning of the pulse. 

\begin{figure}[t!]
\includegraphics[width=\columnwidth]{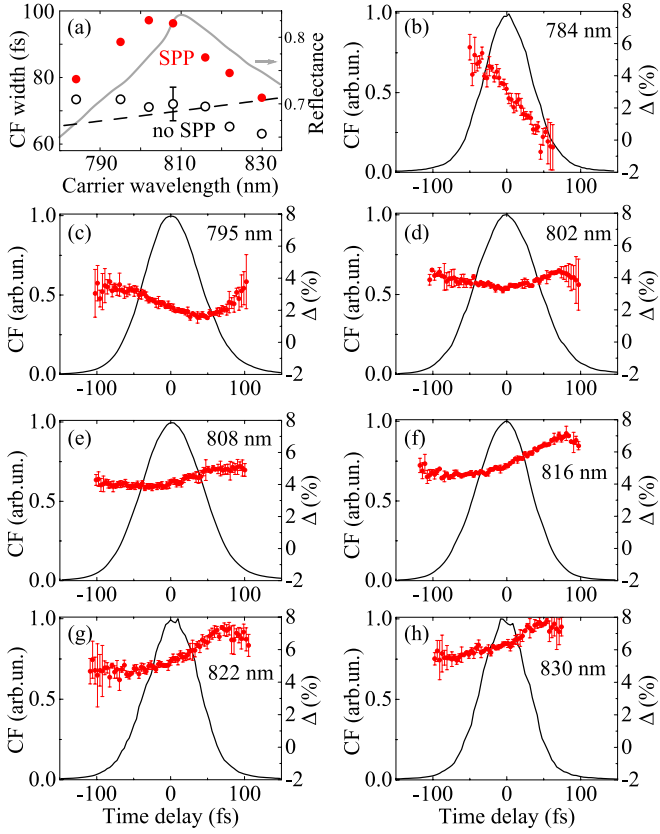}
\caption{(a) Intensity CF width measured with the correlation setup (filled dots) and with an autocorrelator placed in front of the sample (open dots) as a function of the carrier wavelength. The black dashed curve denotes the Fourier-transform limit. The gray curve shows the reflectance spectrum. (b-h) Correlation functions (curves) and the time-dependent TMOKE represented by the $\Delta(\tau)$ function, as denoted in Eq.(\ref{Delta}) (dots with error bars), for the pulses with carrier wavelength values of 784, 795, 802, 808, 816, 822, and 830\,nm, respectively. }\label{fig4}
\end{figure}

We demonstrated the influence of SPPs on the shape of the pulses by measuring the full width at half maximum (FWHM) of the CFs as a function of the carrier wavelength, as presented in Fig.\,\ref{fig4}(a). The CF width spectrum displays good agreement with the SPP resonance, as indicated by the reflection spectrum shown in gray. The open dots denote the width of the intensity autocorrelation function as measured by an autocorrelator placed in front of the sample. The dashed line illustrates the Fourier-transform limit for the sech$^2(t)$-pulses. 


The key results of this work are presented in Fig.\,\ref{fig4}(b-h). A time-dependent magneto-optic effect is demonstrated with the measured $\Delta(\tau)$ function depicted by the red dots superimposed on the corresponding CFs for different $\lambda_c$. The zero time delay is each time defined at the CF maximum. In these experiments, the spectral bandwidth of the pulses was kept at a FWHM of 14\,nm for each $\lambda_c$. The zero dispersion was set to $\lambda_c=800$\,nm, and Fourier-limited pulses were acquired for all of the $\lambda_c$ values in use, thus providing an average time-bandwidth product of $0.31\pm0.1$, as measured by an autocorrelator located in front of the sample. The effect was observed to strongly depend on the part of the SPP resonance that was excited. A gradual increase in $\Delta(\tau)$ was observed for $\lambda_c>802$\,nm. The reverse behavior was observed for $\lambda_c<802$\,nm, where $\Delta(\tau)$ was found to decrease either monotonically or on average. Finally, the TMOKE remained almost constant for $\lambda_c=802$\,nm. Such spectrally selective behavior  provides evidence for the major role of SPPs in determining $\Delta(\tau)$.

The effect was also studied in other experimental configurations. Illuminating the sample with $s$-polarized light should have provided no signal because of the TMOKE properties. This prediction was successfully confirmed, with no detectable signal at the frequency of the external magnetic field under CW illumination. However, a spurious TMOKE signal of  approximately $0.4$\% was obtained for the CF measurement. In this case, the $\Delta(\tau)$ dependence resembled an interference pattern and could be the result of mechanical oscillations in the setup that disturbed the far-field interference pattern formed by the two beams incident to the BBO crystal. Nevertheless, the magnitude of this effect was approximately one order of magnitude lower than that observed with $p$-polarized light and thus does not contradict our main result. The same procedure was also conducted for the sample azimuthally rotated by 90$^\circ$. In this case, the $p$-polarized static TMOKE yielded $(0.2\pm0.02)$\% for all the wavelength values of interest and no pronounced $\Delta(\tau)$ dependence was observed. 

As stated above, the primary effect of the non-trivial $\Delta(\tau)$ dependence can be understood in terms of the magnetization-dependent SPP dispersion relation. The reflection coefficient function of the sample is described by the complex Lorentz spectral line shape
\begin{equation}
r(\omega,M)=\frac{\imath \gamma r(\omega_0)}{\omega_0(M)-\omega+\imath \gamma},\label{resonance}
\end{equation}
where $\omega_0$ is the SPP resonance frequency, $\gamma$ is the inverse SPP lifetime, and $r(\omega_0)$ is the reflection coefficient at the resonance. The central frequency should be magnetization-dependent. Indeed, the phase-matching conditions for the SPP coupled through the --1st diffraction order are as follows:
\begin{equation}
k_{\mbox{\tiny SPP}}(M)-\frac{\omega_0}{c}\sin{\theta}=\frac{2\pi}{d},\label{phasematching}
\end{equation}
where $d$ is the period of the grating. Therefore, Eqs.(\ref{SPP},\ref{phasematching}) establish that $\omega_0$ depends on the magnetic field. Although the explicit expression for $\omega_0(M)$ cannot be obtained, we can estimate the expected wavelength shifts to be approximately 5\,nm, as derived from Eq.\,(\ref{SPP}). The last step before obtaining the expression for $\delta(M,t)$ is to evaluate the impact of $M$ on the reflected pulse envelope function. The latter is found via the convolution theorem:
\begin{eqnarray}
I(M,t)=|E_{\mbox{\tiny out}}(M,t)|^2,\label{intensity}\\
E_{\mbox{\tiny out}}(M,t)=\int\limits_{-\infty}^{\infty}r(M,t^\prime)E_{\mbox{\tiny in}}(t-t^\prime)dt^\prime,
\end{eqnarray}
where $r(M,t)$ is the result of taking the inverse Fourier transform of Eq.(\ref{resonance}):
\begin{equation}
r(M,t)=\sqrt{2\pi}r(\omega_0)\Theta(t)\exp[-\gamma t+\imath \omega_0(M) t].
\end{equation}
Here, $\Theta(t)$ is the Heaviside step function. For the sake of calculation simplicity, let us suppose that the initial pulse is described by a Gaussian envelope:
\begin{equation}
E_{\mbox{\tiny in}}=\exp\left(-\frac{t^2}{2\sigma^2} +\imath \omega_c t\right),\label{Ein}
\end{equation}
with a width of $\sigma$ and a carrier frequency of $\omega_c=2\pi c/\lambda_c$. Eqs.(\ref{intensity}-\ref{Ein}) provide the expression for the reflected pulse:
\begin{eqnarray}
E_{\mbox{\tiny out}}(M,t)= \pi \gamma \sigma r(\omega_0) e^
{-\gamma t - \imath \omega_0 t + \frac{1}{2} \sigma^2 
\left[
\gamma - \imath (\omega_0 - \omega_c)\right]^2}\\ \nonumber
\times\left(1 + 
   \mbox{erf}\left[\frac{t - \sigma^2 (\gamma - \imath (\omega_0 - \omega_c))}{\sqrt2 \sigma}\right]
\right),\label{Eout}
\end{eqnarray}
where erf$(x)$ is the error function and the magnetic field dependence is incorporated in $\omega_0$. The calculated values of $\delta(t)$ and $\Delta(\tau)$ are shown in Fig.\ref{fig3}(b). All the parameters were obtained from the experiment with the exception of $g$, which was varied to fit the modeled $\Delta(\tau)$ dependence to the experimental one. An acceptable qualitative agreement was attained. The data explicitly provide evidence in favor of the aforementioned hypothesis  that the monotonous increase in the TMOKE is due to the greater influence of the SPPs on the tail of the pulse. 

It is peculiar to observe different signs of the $\Delta(\tau)$ time derivative for different $\lambda_c$. We believe there are several reasons for the negative derivative of $\Delta(\tau)$ that causes the Lorentz-resonance interpretation to lose some degree of consistency. Under a 5$^\circ$ incidence angle, the --1st diffraction order is present for $\lambda_c$ values of less than $805$\,nm; therefore, for these wavelength values, spectrally dependent addition imposed by diffraction is expected. However, the shape of the resonance itself cannot be convincingly described to be symmetric. It was previously demonstrated that by considering an asymmetrical Fano-type resonance, one can enrich the variety of pulse shaping options to impose sign-changing on the magnetic field addition to $I(M,t)$ \cite{Vabishchevich2}. 

In conclusion, manifestations of a time-dependent TMOKE were experimentally demonstrated in 45-fs laser pulses reflected from a one-dimensional iron-based magnetoplasmonic crystal. The effect is attributed to exciting SPPs with magnetization-dependent dispersion. The Kerr effect evolution was demonstrated to have either a positive or negative time derivative, depending on the position of the incident pulse's carrier wavelength $\lambda_c$ with respect to the SPP resonance. Proper justification was given for this effect using the Lorentzian spectral line shape approach. Because the iron-based plasmonic crystal studied in this work is a subwavelength-thickness tailorable nanostructure, it is a promising tool for manipulating femtosecond laser pulses with an external magnetic field that may have applications in novel, active, plasmon-based telecom devices.

The authors thank A. A. Grunin and V. V. Rodionova for the sample fabrication and Chang-Won Lee for fruitful discussions. Support from the Russian Foundation for Basic Research and the Ministry of Education and Science of the Russian Federation is acknowledged.

%

\end{document}